\documentclass[conference]{IEEEtran}
\usepackage{cite}
\usepackage[pdftex]{graphicx}
\usepackage[cmex10]{amsmath}
\usepackage{array}
\usepackage[font=footnotesize]{subcaption}
\usepackage[utf8]{inputenc}
\DeclareUnicodeCharacter{A0}{~}
\usepackage[T1]{fontenc}
\usepackage{fixltx2e}
\usepackage{url}
\usepackage{hyperref}
\hypersetup{colorlinks=true, linkcolor=black, citecolor=black, filecolor=black, urlcolor=black, bookmarksdepth=subsection}
\usepackage{booktabs}
\usepackage[autostyle,english=british]{csquotes}
\usepackage{booktabs}
\usepackage{subcaption}
\usepackage{varioref}
\usepackage{algorithm}
\usepackage[noend]{algpseudocode}
\author{\IEEEauthorblockN{Toke Høiland-Jørgensen} \IEEEauthorblockA{Dept. of Computer Science\\ Karlstad University, Sweden\\ toke.hoiland-jorgensen@kau.se} \and \IEEEauthorblockN{Dave Täht} \IEEEauthorblockA{Teklibre\\ Los Gatos, California\\ dave.taht@gmail.com} \and \IEEEauthorblockN{Jonathan Morton} \IEEEauthorblockA{\\Somero, Finland\\chromatix99@gmail.com}\thanks{toke.hoiland-jorgensen@kau.se}}
\date{\today}
\title{Piece of CAKE: A Comprehensive Queue Management Solution for Home Gateways}
\begin{document}

\maketitle
\begin{abstract}
The last several years has seen a renewed interest in smart queue management to
curb excessive network queueing delay, as people have realised the prevalence of
bufferbloat in real networks.

However, for an effective deployment at today's last mile connections, an
improved queueing algorithm is not enough in itself, as often the bottleneck
queue is situated in legacy systems that cannot be upgraded. In addition,
features such as per-user fairness and the ability to de-prioritise background
traffic are often desirable in a home gateway.

In this paper we present Common Applications Kept Enhanced (CAKE), a
\emph{comprehensive network queue management system} designed specifically for home
Internet gateways. CAKE packs several compelling features into an integrated
solution, thus easing deployment. These features include: bandwidth shaping with
overhead compensation for various link layers; reasonable DiffServ handling;
improved flow hashing with both per-flow and per-host queueing fairness; and
filtering of TCP ACKs.

Our evaluation shows that these features offer compelling advantages, and
that CAKE has the potential to significantly improve performance of last-mile
internet connections.
\end{abstract}

\section{Introduction}
\label{sec:intro}
Eliminating bufferbloat has been recognised as an important component in
ensuring acceptable performance of internet connections, especially as
applications and users demand ever lower latencies. The last several years have
established that Active Queue Management and Fairness Queueing are effective
solutions to the bufferbloat problem, and several algorithms have been proposed
and evaluated (e.g.,
\cite{hoiland-jorgensen15:_good_bad_wifi,jarvinen_evaluating_2014,much-ado}).

However, while modern queueing algorithms can effectively control bufferbloat,
effective deployment presents significant challenges. The most immediate
challenge is that the home gateway device is often not directly in control of
the bottleneck link, because queueing persists in drivers or firmware of devices
that cannot be upgraded \cite{hoiland-jorgensen15:_good_bad_wifi}. In addition,
other desirable features in a home networking context (such as per-user
fairness, or the ability to explicitly de-prioritise background applications)
can be challenging to integrate with existing queueing solutions. To improve
upon this situation, we have developed Common Applications Kept Enhanced (CAKE),
which is a \emph{comprehensive network queue management} system designed specifically
for the home router use case.

As outlined below, each of the issues that CAKE is designed to handle has been
addressed separately before. As such, the compelling benefit of CAKE is that it
takes state of the art solutions and integrates them to provide:

\begin{itemize}
\item a high-precision rate-based bandwidth shaper that includes overhead and link layer
compensation features for various link types.
\item a state of the art fairness queueing scheme that simultaneously provides both
host and flow isolation.
\item a Differentiated Services (DiffServ) prioritisation scheme with rate limiting
of high-priority flows and work-conserving bandwidth borrowing behaviour.
\item TCP ACK filtering that increases achievable throughput on highly asymmetrical
links.
\end{itemize}

CAKE is implemented as a \emph{queueing discipline} (qdisc) for the Linux kernel. It
has been deployed as part of the OpenWrt router firmware for the last several
years and is in the process of being submitted for inclusion in the mainline
Linux kernel.\footnote{We include links to the source code, along with the full evaluation
dataset, in an online appendix \cite{hoiland_jorgensen_toke_2018_1226887}.\label{orgc8ef994}}

The rest of this paper describes the design and implementation of CAKE and is
organised as follows: Section \ref{sec:background} outlines the desirable
features of a comprehensive queue management system for a home router, and
recounts related work in this space. Section \ref{sec:cake-algo} describes the
design and implementation of CAKE in more detail, and
Section \ref{sec:perf-eval} evaluates the performance of the various features.
Finally, Section \ref{sec:conclusions} concludes.

\section{Background and Related Work}
\label{sec:background}
As mentioned initially, CAKE is designed to run on a home network gateway. We
have gathered significant experience with implementing such a system in form of
the Smart Queue Management (SQM) system shipped in the OpenWrt router firmware
project, which has guided the design of CAKE.

In this section we provide an overview of the problems CAKE is designed to
address. We are not aware of any previous work addressing the home gateway queue
management challenges as a whole. However, several of the issues that CAKE
addresses have been subject of previous work, and so the following subsections
serve as both an introduction to the design space and an overview of related
work.

The four problems we seek to address are bandwidth shaping, queue management and
fairness, DiffServ handling and TCP ACK filtering. These are each treated in
turn in the following sections.

\subsection{Bandwidth Shaping}
\label{sec:background-shaper}
A queue management algorithm is only effective if it is in control of the
bottleneck queue. Thus, queueing in lower layers needs to be eliminated, which
has been achieved in Linux for Ethernet \cite{bql} and WiFi
\cite{ending-the-anomaly}. However, eliminating queueing at the link layer is not
always possible, either because the driver source code is unavailable, or
because the link-layer is implemented in inaccessible hardware or firmware
(either on the same device or a separate device, such as a DSL modem).

As an alternative, queueing in the lower layers can be avoided by deploying a
bandwidth shaper as part of the queue management system. By limiting the traffic
traversing the bottleneck link to a bandwidth that is slightly less than the
physical capacity of the link itself, queueing at the physical bottleneck can be
eliminated and bufferbloat avoided. Such bandwidth shaping can be performed by a
token bucket-based shaper (as is well-known from ATM networks, e.g.,
\cite{niestegge1990leaky}), or by a rate-based shaper (which is known from video
streaming applications, e.g., \cite{eleftheriadis1995constrained}).

The use of a shaper to move the link bottleneck wastes the bandwidth that is the
difference between the actual physical link capacity, and the set-point of the
shaper. To limit this waste, the shaper needs to be set as close to the actual
link bandwidth as possible, while avoiding sending bursts of packets at a rate
that is higher than the actual capacity. To achieve this, accurate timing
information on a per-packet basis is needed. In addition, the shaper must
account for link-layer framing and overhead. For instance, DSL links using ATM
framing split up data packets into an integer number of fixed-size cells, which
means that the framing overhead is a step function of packet size, rather than a
fixed value.

\subsection{Queue Management}
\label{sec:background-queueing}
Having control of the bottleneck queue makes it possible to implement effective
queue management that can all but eliminate bufferbloat. Such a queue management
scheme usually takes the form of an Active Queue Management (AQM) algorithm,
combined with a form of fairness queueing (FQ). Several such schemes
exist, and extensive evaluation is available in the literature (e.g.,
\cite{hoiland-jorgensen15:_good_bad_wifi,jarvinen_evaluating_2014,much-ado,rao_analysis_2014,aqm-survey,benameur2013latency}).

Among the state of the art algorithms in modern queue management, is the
FQ-CoDel algorithm \cite{rfc8290}. FQ-CoDel implements a hybrid AQM/fairness
queueing scheme which isolates flows using a hashing scheme and schedules them
using a Deficit Round-Robin (DRR) \cite{drr} scheduler. In addition, FQ-CoDel
contains an optimisation that provides implicit service differentiation for
sparse (low-bandwidth) flows, similar to \cite{kortebi_implicit_2005,abdesselem_kortebi_cross-protect:_2004}. Evaluations of FQ-CoDel have shown that it
achieves low queueing latency and high utilisation under a variety of scenarios
\cite{hoiland-jorgensen15:_good_bad_wifi,much-ado}.

However, while the FQ-CoDel scheduler provides flow isolation and fairness, the
transport layer flow is not always the right level of fairness in the home
gateway use case. Often, additional isolation between \emph{hosts} on the network is
desirable; and indeed this per-host isolation was the most requested
feature of the SQM system. Host isolation is straight-forward to implement in
place of flow fairness in any fairness queueing based scheme (by simply changing
the function that maps packets into different queues), but we are not aware of
any practical schemes prior to CAKE that implement \emph{both} host and flow fairness.

\subsection{DiffServ Handling}
\label{sec:background-diffserv}
Even though flow-based fairness queueing offers a large degree of separation
between traffic flows, it can still be desirable to explicitly treat some
traffic as higher priority, and to have the ability to mark other traffic as low
priority. Since a home network generally does not feature any admission control,
any prioritisation scheme needs to be robust against attempts at abuse (so,
e.g., a strict priority queue does not work well). In addition, enabling
prioritisation should not affect the total available bandwidth in the absence of
marked traffic, as that is likely to cause users to turn the feature off.

Prioritisation of different traffic classes can be performed by reacting to
DiffServ markings \cite{rfc4594}. This is commonly used in WiFi networks, where
DiffServ code points map traffic into four priority levels \cite{rfc8325}. For the
home gateway use case, various schemes have been proposed in the literature
(e.g., \cite{hwang2005qos}), but as far as we are aware, none have seen
significant deployment.

\subsection{TCP ACK Filtering}
\label{sec:background-ack-filtering}
TCP ACK filtering is an optimisation that has seen some popularity in highly
asymmetrical networks \cite{wu1999ack}, and especially in cable modem deployments
\cite{storfer2003enhancing}. The technique involves filtering (or \emph{thinning}) TCP
acknowledgement (ACK) packets by inspecting queues and dropping ACKs if a TCP
flow has several consecutive ACKs queued. This can improve performance on highly
asymmetrical links, where the reverse path does not have sufficient capacity to
transport the ACKs produced by the forward path TCP flow. However, ACK filtering
can also have detrimental effects on performance, for instance due to cross
layer interactions \cite{kim2006cross}.

\section{The Design of CAKE}
\label{sec:cake-algo}
The design of CAKE builds upon the basic fairness scheduler design of FQ-CoDel,
but adds features to tackle the areas outlined in the previous section. The
following sections outline how CAKE implements each of these features.

\subsection{Bandwidth Shaping}
\label{sec:cake-shaper}
CAKE implements a rate-based shaper, which works by scheduling packet
transmission at precise intervals using a virtual transmission clock. The clock
is initialised by the first packet to arrive at an empty queue, and thereafter
is incremented by the calculated serialisation delay of each transmitted packet.
Packets are delayed until the system time has caught up with the virtual clock.
If the clock schedule is reached while the queue is empty, the clock is reset
and the link goes idle.

This shaper handles bandwidth ranging over several orders of magnitude, from
several Kbps to several Gbps. In addition, the rate-based shaper does not
require a burst parameter, which simplifies configuration as compared to a
token-bucket shaper. It also eliminates the initial burst observed from
token-bucket shapers after an idle period. This is important for controlling the
bottleneck queue, as this initial burst would result in queueing at the real
bottleneck link.

\subsubsection{Overhead and Framing Compensation}
\label{sec:overhead}
As mentioned in Section \ref{sec:background-shaper} above, the shaper accounts
for the actual size of a packet on the wire, including any encapsulation and
overhead, which allows the rate to be set closer to the actual bottleneck
bandwidth, thus eliminating waste. We believe it is safe to set a rate within
0.1\% of the actual link rate when the overhead compensation is configured
correctly, with a margin mainly required to accommodate slight variations in the
actual bottleneck link bandwidth, caused by, e.g., clock drift in the hardware.

CAKE implements an overhead compensation algorithm which begins by determining
the size of the network-layer packet, stripped of any MAC layer encapsulation.
Having determined the network-layer packet size, the configured overhead can be
added to yield the correct on-the-wire packet size, followed optionally by a
specialised adjustment for ATM or PTM framing. This algorithm is shown in
Algorithm \ref{alg:shaper}.

Using the network-layer packet size and adding a manually configured overhead
value is required because the values reported by the kernel are often wrong due
to idiosyncrasies of the CPE unit. While this does make configuration a bit more
complex, we seek to alleviate this by providing keywords for commonly used
configurations.

As part of the overhead compensation, CAKE also optionally splits "super
packets" generated by hardware offload features. These super packets are
essential for operating at high bandwidths, as they help amortise fixed network
stack costs over several packets. However, at lower bandwidths they can hurt
latency, in particular when a link with a high physical bandwidth is shaped to a
lower rate. For this reason, we conditionally split super packets when shaping
at rates lower than 1 Gbps. This allows CAKE to ensure low latency at lower
rates, while still scaling to full line rate on a 40Gbps link.

\begin{algorithm}[t]
\caption{Shaping and overhead compensation algorithm. \small \emph{T\_next} is the time at which the next packet is eligible for tranmission.}
\label{alg:shaper}
\begin{algorithmic}[1]
\small
\Function{enqueue}{\emph{pkt}}
  \State $\textit{net\_len} \gets \textit{pkt.len}\, -\, $\Call{network\_offset}{\textit{pkt}}
  \State $\textit{adj\_len} \gets \textit{net\_len} + \textit{overhead}$
  \If {ATM framing is enabled}
    \State $\textit{adj\_len} \gets$ \Call{ceiling}{\textit{adj\_len} / 48} * 53
  \ElsIf {PTM framing is enabled}
    \State $\textit{adj\_len} \gets$ \Call{ceiling}{\textit{adj\_len} / 64} * 65
  \EndIf
  \State $\textit{pkt.adj\_len} \gets \textit{adj\_len}$
  \If {\emph{backlog} is zero and \emph{T\_next} is after \emph{Now}}
    \State $\textit{T\_next} \gets \textit{Now}$
  \EndIf
\EndFunction

\Function{dequeue}{}
  \If {\emph{T\_next} is after \emph{Now}}
    \State Schedule interrupt at \textit{T\_next}
    \State \Return Nil
  \EndIf
  \State $\textit{pkt} \gets$ Choose Packet
  \State $\textit{T\_next} \gets \textit{T\_next} + \textit{pkt.adj\_len} * \textit{time\_per\_byte}$
  \State \Return \textit{pkt}
\EndFunction
\end{algorithmic}
\end{algorithm}
\subsection{Flow Isolation and Hashing}
\label{sec:flow-isolation}
\begin{figure}[htbp]
\centering
\includegraphics[width=\linewidth]{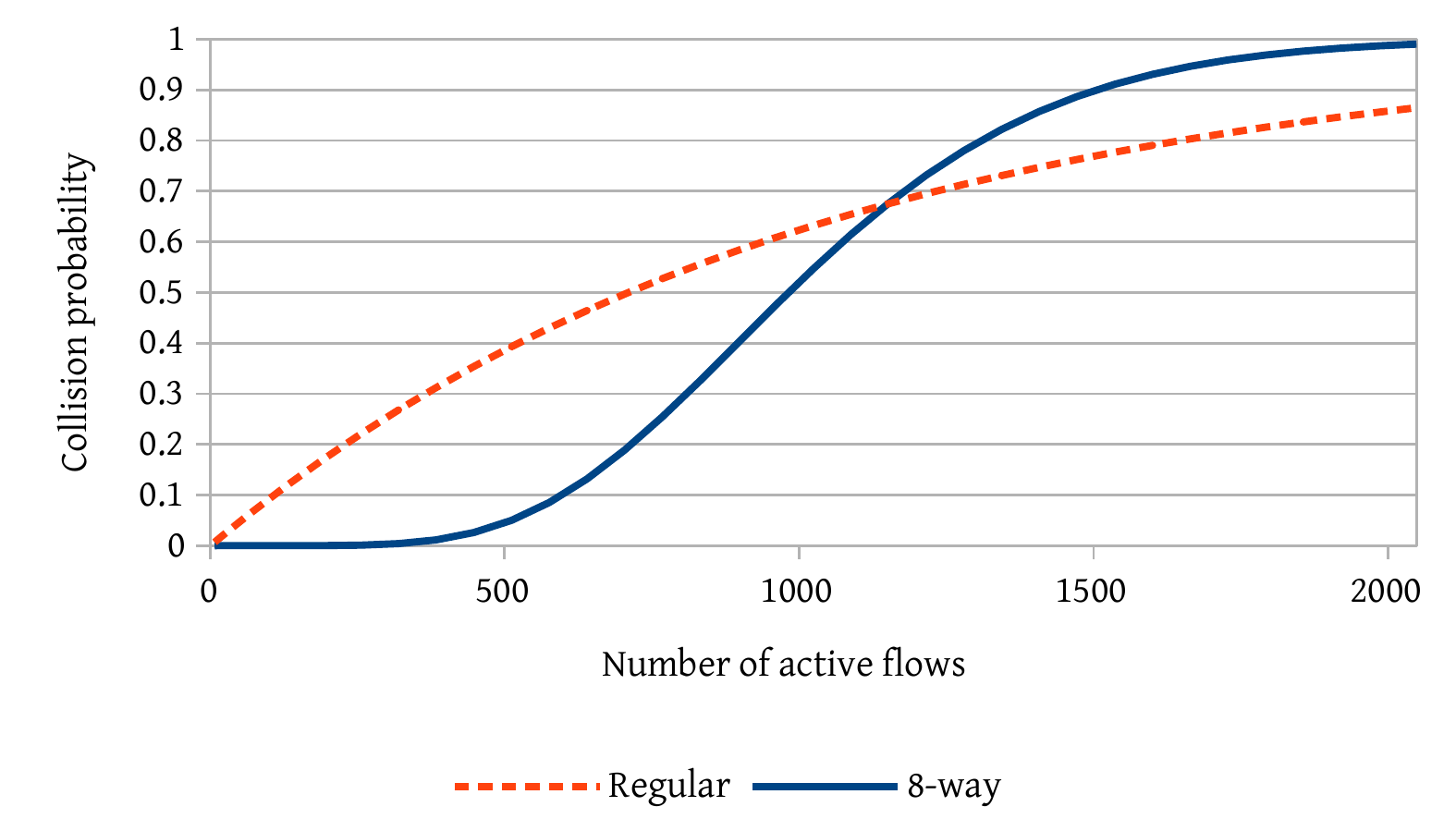}
\caption{\label{fig:hash-coll}
Probability that a new flow will experience a hash collision, as a function of the number of active flows. 1024 total queues.}
\end{figure}

CAKE replaces the direct hash function used in FQ-CoDel with an 8-way
set-associative hash. While set-associative hashing has been well-known for
decades as a means to improve the performance of CPU caches
\cite{smith1978comparative}, it has not seen much use in packet scheduling.
Conceptually, a \(k\mathrm{-way}\) set-associative hash with \(n\) total buckets can
be thought of as a plain hash with \(n/k\) buckets that is only considered to have
a collision if more than \(k\) items hash into the same bucket. As can be seen in
Figure \ref{fig:hash-coll}, this significantly reduces the hash collision probability
up to the point where the number of flows is larger than the number of
queues.\footnote{See how we computed these probabilities in the online appendix.}

\subsubsection{Host Isolation}
\label{sec:host-isolation}
With flow fairness, hosts can increase their share of the available bandwidth by
splitting their traffic over multiple flows. This can be prevented by providing
host fairness at the endpoint IP address level, which CAKE can do in addition to
flow fairness.

The host isolation is simple in concept: The effective DRR quantum is divided by
the number of flows active for the flow endpoint. This mechanism can be
activated in three different modes: source address fairness, in which hosts on
the local LAN receive equal share, destination address fairness, in which
servers on the public internet receive an equal share, or "triple isolate" mode,
in which the maximum of the source and destination scaling is applied to each
flow. CAKE also hooks into the Linux kernel Network Address Translation (NAT)
subsystem to obtain the internal host address of a packet, which would otherwise
be obscured since packets are queued after NAT is applied.

\begin{algorithm}[t]
\caption{\small Host isolation algorithm.}
\label{alg:host-isolation}
\begin{algorithmic}[1]
\small
\Function{enqueue}{\emph{pkt}} \label{ln:beg-enq}
  \State $\textit{flow\_hash} \gets$ \Call{hash}{\textit{pkt.hdr}}
  \State $\textit{src\_hash} \gets$ \Call{hash}{\textit{pkt.src\_ip}}
  \State $\textit{dst\_hash} \gets$ \Call{hash}{\textit{pkt.dst\_ip}}
  \State $\textit{flow} \gets \textit{flows[flow\_hash]}$
  \If {\emph{flow} is not active}
    \State $\textit{hosts[src\_hash].refcnt\_src++}$
    \State $\textit{hosts[dst\_hash].refcnt\_dst++}$
    \State $\textit{flow.active} \gets 1$
    \State $\textit{flow.src\_id} \gets \textit{src\_hash}$
    \State $\textit{flow.dst\_id} \gets \textit{dst\_hash}$
  \EndIf \label{ln:end-chk-st}
\EndFunction \label{ln:end-enq}

\Function{get\_quantum}{\emph{flow}}
  \State $\textit{refcnt\_src} \gets \textit{hosts[flow.src\_id].refcnt\_src}$
  \State $\textit{refcnt\_dst} \gets \textit{hosts[flow.dst\_id].refcnt\_dst}$
  \State $\textit{host\_load} \gets$ \Call{max}{\textit{refcnt\_src}, \textit{refcnt\_dst}, 1}
  \State \Return $\textit{flow.quantum} / \textit{host\_load}$
\EndFunction
\end{algorithmic}
\end{algorithm}

CAKE accomplishes this scaling as shown in Algorithm \ref{alg:host-isolation}:
When a packet is enqueued it is hashed into a queue using the transport layer
port numbers along with the source and destination IP addresses. In addition,
two separate hashes are performed on the packet destination IP address and
source IP address. A separate set of hash buckets is kept for these address
hashes. These buckets do not contain a queue of packets, but instead a data
structure that keeps two reference counts for each IP address, which track the
number of active flows with the given address as source and destination,
respectively.

The per-IP reference counts are used to modify the quantum for each active flow.
When a flow is scheduled, its "host load" is calculated as the maximum of the
reference counts for its source and destination IP addresses. The effective
quantum of the flow is simply divided by this load value, which achieves the
desired scaling.

\subsection{DiffServ handling}
\label{sec:cake-diffserv}
CAKE provides a small number of preset configurations, which map each DiffServ
code point into a priority tier. If the shaper is in use, each priority tier
gets its own virtual clock, which limits that tier's rate to a fraction of the
overall shaped rate. When dequeueing a packet, the algorithm simply picks the
highest-priority tier which both has queued traffic and whose schedule is due,
if one exists. To allow tiers to borrow excess bandwidth from one another, the
dequeue algorithm also tracks the earliest schedule time of all non-empty tiers,
and if no other eligible tier is available, that tier is picked instead (within
the overall shaper limits).

When the shaper is not in use, CAKE instead uses a simple weighted DRR mechanism
to schedule the different priority tiers, with the same weights as the shaper
fractions mentioned above. This has weaker precedence guarantees for
high-priority traffic, but provides the same proportional capacity reservation
and the ability to borrow spare capacity from less than fully loaded tiers.

CAKE defaults to a simple, three-tier mode that interprets most code points as
"best effort", but places CS1 traffic into a low-priority "bulk" tier which is
assigned \(1/16\) of the total rate, and a few code points indicating
latency-sensitive or control traffic (specifically TOS4, VA, EF, CS6, CS7) into
a "latency sensitive" high-priority tier, which is assigned 1/4 rate. The other
DiffServ modes supported by CAKE are a 4-tier mode matching the 802.11e
precedence rules \cite{rfc8325}, as well as two 8-tier modes, one of which
implements strict precedence of the eight priority levels.

\subsection{ACK filtering}
\label{sec:ack-filtering}
CAKE contains an ACK filtering mechanism that drops \emph{redundant} ACKs from a TCP
flow. The mechanism takes advantage of the per-flow queueing by scanning the
queue after every packet enqueue, to identify a pure ACK (i.e., an ACK with no
data) that was made redundant by the newly enqueued packet. An ACK is only
filtered if the newly enqueued packet contains an acknowledgement of \emph{strictly
more} bytes than the one being filtered. In particular, this means that
duplicate ACKs are not filtered, so TCP's fast retransmit mechanism is not
affected. In addition, the filter parses TCP headers and only drops a packet if
that will not result in loss of information at the sender; and packets with
unknown headers are never dropped, to avoid breaking future TCP extensions. The
filter has two modes of operation: a conservative mode that will always keep at
least two redundant ACKs queued, and an aggressive mode, that only keeps the
most recently enqueued ACK.

\section{Performance Evaluation}
\label{sec:perf-eval}
In this section, we present a performance evaluation of CAKE. All tests are
performed on a testbed that emulates a pair of hosts communicating through a
low-bandwidth link. We use the Flent testing tool \cite{flent} to run the tests,
and the data files are available on the companion web site.\textsuperscript{\ref{orgc8ef994}} Unless otherwise
stated below, all tests are run on a symmetrical 10 Mbps link with 50 ms
baseline latency. Our basic test is the Real-Time Response Under Load test,
which consists of running four TCP flows in each traffic direction, along with
three different latency measurement flows \cite{rrul}.

\begin{figure}[htbp]
\centering
\includegraphics[width=\linewidth]{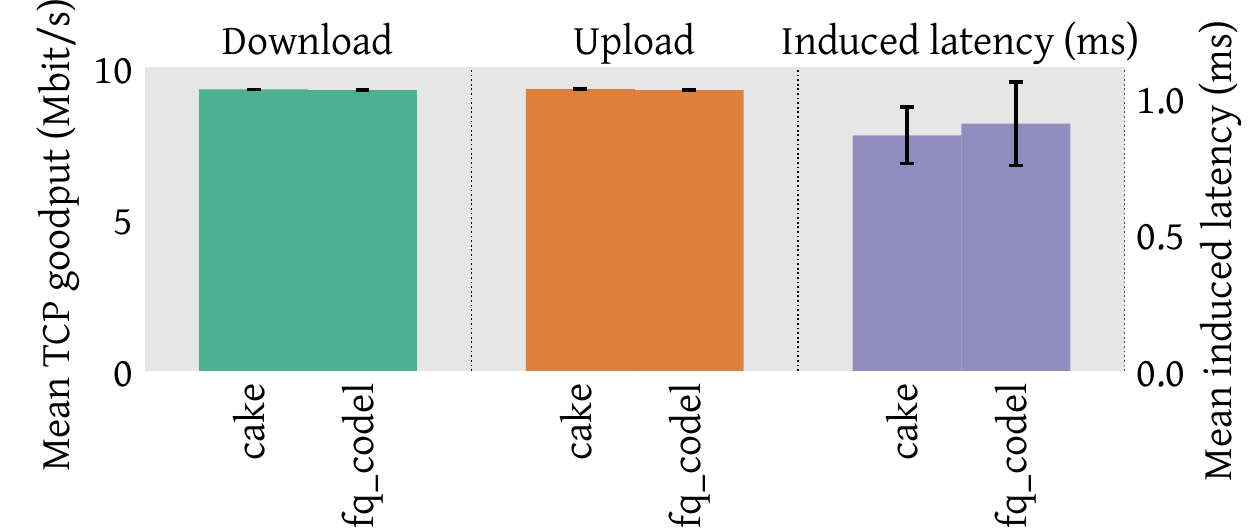}
\caption{\label{fig:rrul-cake-fq-codel}
Baseline throughput and latency of CAKE and FQ-CoDel on a 10 Mbps link.}
\end{figure}

As can be seen in Figure \ref{fig:rrul-cake-fq-codel}, the baseline performance of
CAKE is comparable to that of FQ-CoDel: both achieve low latency and high
throughput in the baseline test. This is expected, since CAKE is derived from
FQ-CoDel. For a more comprehensive comparison of FQ-CoDel with other queue
management algorithms, we refer the reader to
\cite{hoiland-jorgensen15:_good_bad_wifi}. Instead, the remainder of this
evaluation focuses on the features outlined in the previous sections.

\subsection{Host Isolation}
\label{sec:eval-host-isolation}
To evaluate the host isolation feature of CAKE, we run a varying number of TCP
flows between two source hosts and four destination hosts. Source host A runs
one flow to each of destination hosts A and B, and two flows to destination host
C, while source host B runs one flow to each of destination hosts C and D. This
makes it possible to demonstrate the various working modes of CAKE's host
isolation feature.

\begin{figure}[htbp]
\centering
\includegraphics[width=\linewidth]{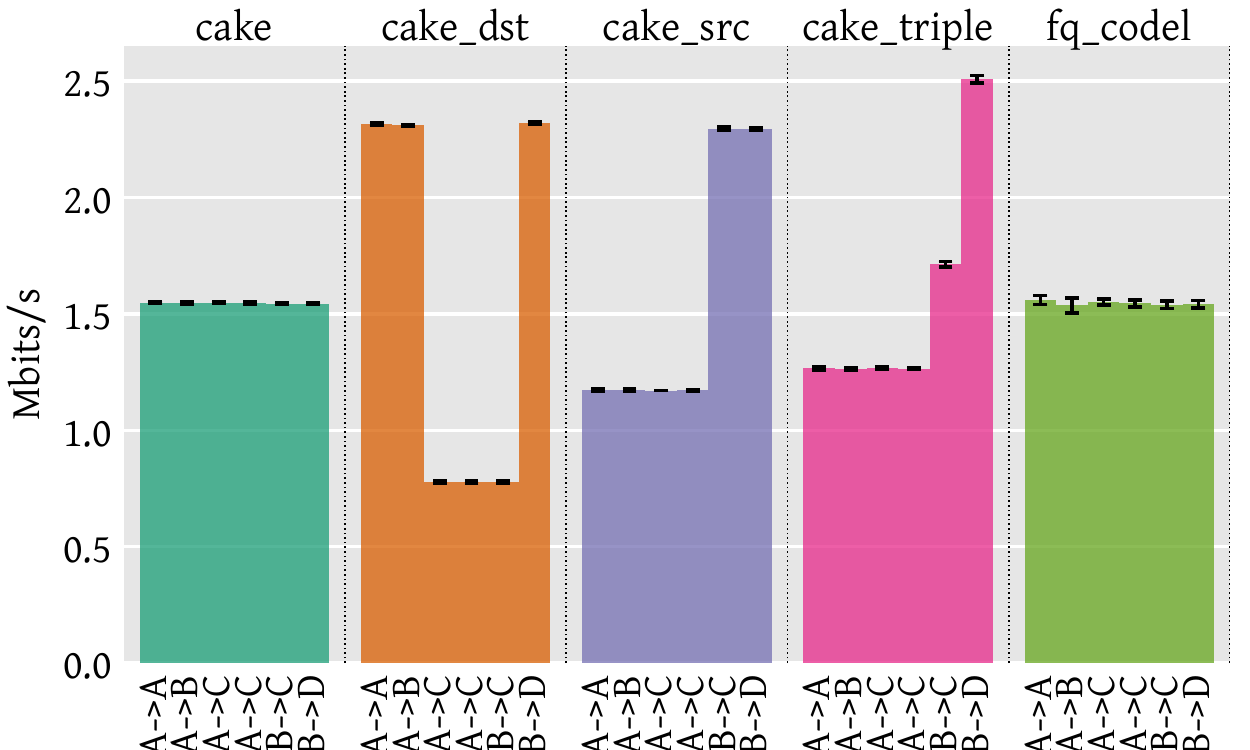}
\caption{\label{fig:triple-isolate}
Host isolation performance with TCP flows from two source hosts to four destination hosts. The columns show different algorithms;  each bar shows the average flow goodput.}
\end{figure}

The result of this test is shown in Figure \ref{fig:triple-isolate}. It shows four
configurations of CAKE (no host isolation, source host isolation, destination
host isolation and triple isolation) and a test with FQ-CoDel as the queue
management algorithm. As can be seen in the figure, both FQ-CoDel and CAKE with
no host isolation provide complete fairness between all six flows.

The figure also clearly shows the various modes of flow isolation supported by
CAKE: In destination fairness mode (second column), the four destination hosts
get the same total share, which results in each of the three flows to
destination host C getting \(1/3\) of the bandwidth of the three other hosts
(which only have one flow each). Similarly, in source fairness mode (third
column), the two source hosts share the available capacity, which results in the
two flows from source B getting twice the share each compared to the four flows
from host A.

In the triple isolation case, we see the flow bandwidths correspond to the
quantum scaling outlined in Algorithm \ref{alg:host-isolation}: The first four
flows get their quantum scaled by \(1/4\) since there are four flows active from
host A. The fifth flow gets its quantum scaled by \(1/3\) since there are three
flows active to host C. And finally, the last flow gets its quantum scaled by
\(1/2\) as there are two flows active from host B.

\subsection{DiffServ Handling}
\label{sec:eval-diffserv}
To demonstrate the DiffServ prioritisation features of CAKE we perform two
tests: An RRUL test with each flow marked with a different DiffServ priority,
and another test where a high-priority fixed-rate flow competes with several TCP
flows.

\begin{figure}[htbp]
\centering
\includegraphics[width=\linewidth]{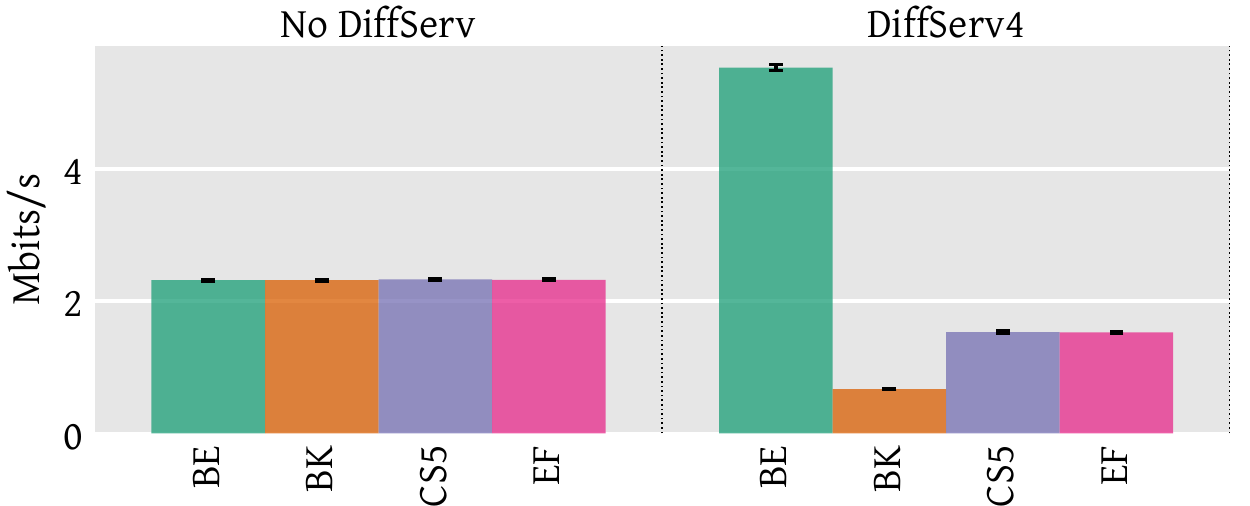}
\caption{\label{fig:diffserv-tcp}
TCP flows on different DiffServ code points.}
\end{figure}

The result of the former test is seen in Figure \ref{fig:diffserv-tcp}. This shows
that when DiffServ mode is not enabled, all four flows get the same share of the
available bandwidth, while in the DiffServ-enabled case, the Best Effort (BE)
flow gets most of the bandwidth. This latter effect is important for two
reasons: First, it shows that a flow marked as background (BK) is successfully
de-prioritised and gets less bandwidth. Secondly, it shows that the
high-priority flows (CS5 and EF) are limited so as to not use more than the
share of the bandwidth allocated to the high-priority DiffServ classes.

\begin{figure}[htbp]
\centering
\includegraphics[width=\linewidth]{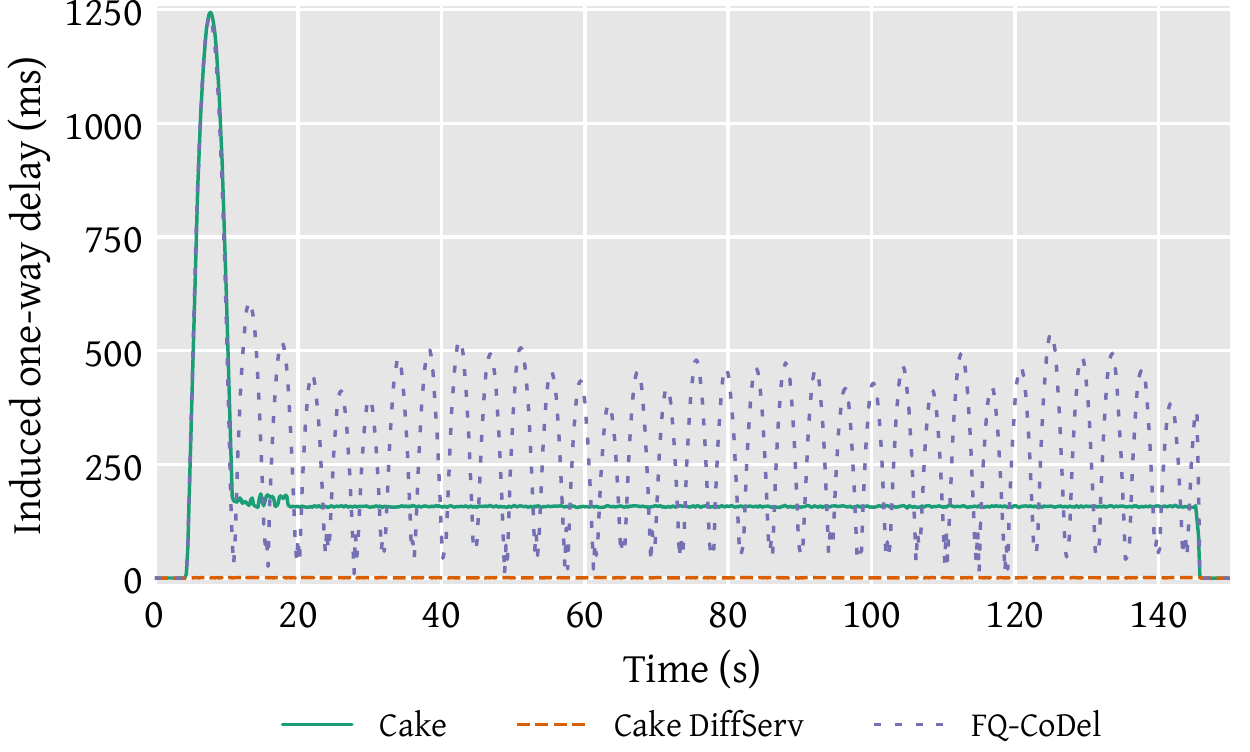}
\caption{\label{fig:diffserv-voip}
Latency over time of a 2 Mbps fixed-rate flow with 32 competing bulk flows on a 10 Mbps link. The Y-axis shows additional latency above the base latency of 50 ms. The bulk flows start after 5 seconds.}
\end{figure}

To look at the latency performance of a high-priority flow, we turn to
Figure \ref{fig:diffserv-voip}. This shows the latency over time of a fixed-rate 2
Mbps flow, which marks its packets with the high-priority EF DiffServ marking.
This is meant to represent a real-time video conversation. In the test,
the flow competes with 32 bulk TCP flows. As can be seen in the figure, both
FQ-CoDel and CAKE with DiffServ prioritisation disabled fail to ensure low
latency for the high-priority flow. Instead, when the bulk flows start after
five seconds, a large latency spike is seen, since the real-time flow has to
wait for the initial packets of the 32 TCP flows. This causes the real-time flow
to build a large queue for itself (since it does not respond to congestion
signals), which then drains slowly back to a steady state around 200 ms (for
CAKE) or oscillating between 50 and 500 ms (for FQ-CoDel). In contrast, the
DiffServ-enabled CAKE keeps the real-time flow completely isolated from the bulk
TCP flows, ensuring it sees no added latency over the duration of the test.

\subsection{ACK Filtering}
\label{sec:eval-ack-filtering}
\begin{figure}[htbp]
\centering
\includegraphics[width=\linewidth]{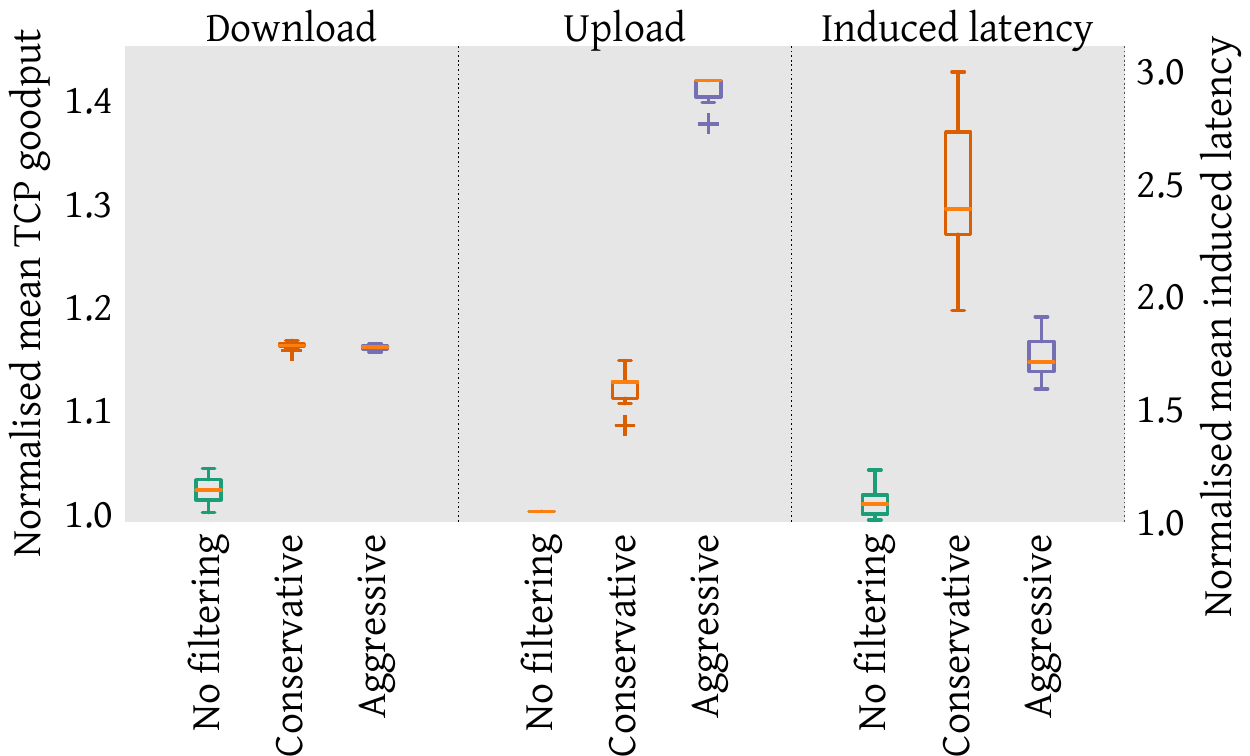}
\caption{\label{fig:ack-filtering}
ACK filtering performance on a 30/1 Mbps link. The graph scales are normalised to the "No filtering" case. The download and upload value ranges are 24.5-27.5 Mbps and 0.45-0.7 Mbps, respectively. The latency range is 2.6-7.5 ms.}
\end{figure}

Figure \ref{fig:ack-filtering} shows the performance of ACK filtering on a highly
asymmetrical link with 30 Mbps download capacity and only 1 Mbps upload
capacity. On this link, we run four simultaneous TCP uploads and four
simultaneous TCP downloads. The results of this are shown in
Figure \ref{fig:ack-filtering}, which shows the aggregate throughput of all four flows
in each direction, along with the added latency of a separate measurement flow.
Values are normalised to the baseline without ACK filtering to be able to fit on
a single graph. As the figure shows, we see a goodput improvement of around 15\%
in the downstream direction caused by either type of ACK filtering, which shows
that insufficient bandwidth for ACKs can impact transfers in the other
direction. For upload, the conservative filtering increases goodput by about
10\%, while the aggressive filtering increases throughput by as much as 40\%,
simply by reducing the bandwidth taken up by ACK packets. We attribute the
increase in latency to increased congestion in the downlink direction, which is
alleviated somewhat by fewer ACKs being queued in the upstream direction in the
aggressive case. The absolute magnitude of the latency increase is only 5 ms.

\section{Conclusions}
\label{sec:conclusions}
CAKE is a comprehensive queue management system for home gateways, that packs
several compelling features into an integrated solution, with reasonable
defaults to ease configuration. These features include: bandwidth shaping with
overhead compensation for various link layers; reasonable DiffServ handling;
improved flow hashing with both per-flow and per-host queueing fairness; and
filtering of TCP ACKs.

Our evaluation shows that these features offer compelling advantages, and we
believe CAKE has the potential to significantly improve the performance of
last-mile internet connections. CAKE is open source and ready for deployment,
and already ships in the OpenWrt router firmware distribution.

\section*{Acknowledgements}
\addcontentsline{toc}{section}{Acknowledgements}

The authors would like to thank the Bufferbloat and OpenWrt communities for
their work on the implementation and testing of CAKE. In particular, Kevin
Darbyshire-Bryant was instrumental in enabling NAT-awareness, Ryan Mounce
contributed the original ACK filtering code, Sebastian Moeller helped get the
overhead compensation right and Anil Agarwal helped with the hash collision
probability calculations.

\bibliographystyle{IEEEtran}
\bibliography{cake}
\end{document}